\documentclass{ws-p8-50x6-00}
\usepackage{graphicx}
\begin{document}

\title{On the Critical Mass: the case of white dwarfs}

\author{Remo J. Ruffini}

\address{Physics Department, University of Rome, La Sapienza,
00185 Rome, Italy}  

\address{I.C.R.A.-International Center for Relativistic
Astrophysics c/o Physics Department, University of Rome, La Sapienza,
00185 Rome, Italy\\E-mail: ruffini@icra.it}

\maketitle

\abstracts{We recall the generalization of the Feynman-Metropolis-Teller approximation for a compressed atom using a relativistic Fermi-Thomas model. These results within a Wigner-Seitz approximation lead to a new equation of state for white dwarfs and to a new value of their critical mass, smaller than the one obtained by Chandrasekhar. The possible observations of these effects in binary neutron stars are outlined.}

\section{Introduction}

The problem of critical mass against gravitational collapse is one of the most important in relativistic astrophysics and has been the subject of a very large number of theoretical investigations with many observational consequences, the most important one being the paradigm for the identification of black holes in binary X-ray sources. The concept of critical mass appears in a large variety of astrophysical settings: white dwarfs, neutron stars, galactic nuclei. The occurrence of gravitational collapse in galactic nuclei  is still matter of great debate  both from the observational and theoretical point of view: the physical conditions leading to the onset of gravitational collapse, even the matter constituents of such a process, are still far from settled although the existence of extremely massive relativistic systems in the range of $10^6-10^8 M_\odot$ in galactic nuclei is certain. The  occurrence of gravitational collapse in neutron stars has obtained grandiose observational verification in binary X-ray sources although, from a theoretical point, it still presents problems of unsurpassed difficulty due to theoretical and observational unknowns in the behavior of matter compressed at supra-nuclear densities. In this case only a firm  upper limit on the numerical value of the critical mass has been obtained from first principles: the validity of general relativity, the causality non-violation principle and the behavior of matter at a fiducial nuclear density.\cite{rhoades}

In this respect the case of white dwarfs appears most interesting: it is amenable to a complete theoretical treatment which is also testable  by refined observational measurements thanks to the superb observations of binary pulsars. The critical mass for white dwarfs occurs at densities relatively low in comparison with nuclear densities and many of the intricacies of the behavior of matter in bulk at supra-nuclear densities can be completely neglected. On the other hand the theoretical analysis of the white dwarf critical mass can become the arena for a detailed treatment of gravitational and electromagnetic effects at relativistic regimes  in the fully tested region of validity of these field theories: general relativity, quantum theory and electromagnetism. I will devote this contribution to  this topic which has reached complete maturity and in principle can give important observational clues for the understanding of astrophysical high-precision measurements.

\section{Historical Background}

That white dwarfs cannot have an arbitrary large mass and that a critical mass against gravitational collapse must exist in their configuration of equilibrium reachable by increasing their central densities was clearly understood in the thirties independently by Subranaman Chandrasekhar\cite{CH1,CH2} and by Lev Davidovich Landau.\cite{L} Their treatments differ enormously in style.

After publishing his result in an article, Chandrasekhar gave his derivation in a 343 page classic book\cite{chandra} which summarized: a) some of the results originally presented in  German in the classic textbook {\it Gaskugeln} by Emden, b) basic aspects of quantum mechanics, developed in the classic book of R.H. Fowler\cite{fowler} and in the work of P.A.M. Dirac, and 3)  the detailed computations of the mass-radius relation for white dwarfs and their agreement with observations with the clear statement of the existence of a critical mass equal to  $1.45M_\odot$.

The work of Lev Landau is totally different in its essential approach: Landau also uses all the expertise which could be gained from the classic work of Emden and quantum theory, at that time just formulated, and the corresponding relativistic corrections. Similarly he reaches the conclusion of the  existence of a critical mass for a white dwarf, but now, everything, in a dramatic paper of no more then {\em three} pages. There by the stroke of crucial shortcuts in the physics of self-gravitating degenerate matter in  relativistic regimes, Landau reaches  that unexpected  conclusion. Landau's paper is not only beautiful for the masterful interplay of physics and astronomy, but can also be considered the first paper introducing a new style of work in the scientific analysis of astronomical phenomena.

Another non-negligible difference must be mentioned distinguishing the work of Landau and Chandrasekhar: Landau {\em did not} believe the possible existence in nature of the new concept of a critical mass he had just obtained, while Chandrasekhar never revealed such a critical attitude towards it.

At the time of presentation of his work in Cambridge, Chandra met with very strong resistance and criticism from Arthur Eddington, so much so that his paper published in the proceedings of the Royal Astronomical Society  was preceded by an article of strong criticism by Eddington himself which clearly stated, with reference to Chandra's work, ``... but its physical foundation does not inspire confidence, since it is a combination of relativistic mechanics with non-relativistic quantum theory."

I was always interested in understanding some of the reasons which motivated Eddington in his criticism. After all, at the time, Eddington was one of the greatest experts in the theory of general relativity and was the author of fundamental textbooks both in astronomy\cite{E1} and general relativity:\cite{E2} for me more than his expressed criticism toward Chandrasekhar's work, it was interesting to understand whether there was any fundamental theoretical issue adopted by Chandrasekhar which could be disturbing to the eyes of a relativist. 

The answer to Eddington's criticism came finally in 1974 from Chandrasekhar himself in his outstanding  article ``Why are the stars as they are" presented at the Varenna Summer school on the Physics and Astrophysics of Neutron Stars and Black Holes.\cite{giacco}

Nevertheless reconsidering recently the issue of the white dwarf critical mass, I have found a basic unproven assumption in Chandrasekhar's work. The distribution of the degenerate electrons is assumed to be uniform in the stars and unaffected by Coulomb interaction with the nuclei all the way to infinite densities. Indeed in Chandrasekhar's work the value of the critical mass is reached for an infinite central density of the star.
On this issue Landau had made two important remarks:
\begin{enumerate}
\item In 1938, following the earlier suggestions by Hund,\cite{hund} Landau\cite{Dau} pointed out {\it ``it is well known that matter consists of nuclei and electrons. Nevertheless it can be shown that in bodies of very large mass, this usual ``electronic'' state of matter can become unstable. The reason for this lies in the fact that the ``electronic'' state of matter does not lead to extremely great densities, because at such densities electrons form a Fermi gas having an immense pressure. On the other hand, it is easy to see that matter can go into another state which is much more compressible---the state where all the nuclei and electrons have combined to form neutrons.''}. Later Yakov Borisovich Zel'dovich\cite{YBZ} pointed out, quantitatively, that this ``neutron condensation'' for inverse beta decay starts to occur for hydrogen already at $10^7 \frac{g}{cm^3}$. Then Harrison et al\cite{wheeler} pointed out that all configurations of white dwarfs after the onset of the inverse beta decay are indeed unstable and the critical mass is reached not at an infinite density but at precisely the finite density marked by the onset of inverse beta decay. This result was confirmed by the analysis of pulsational modes of White Dwarfs by Chandrasekhar \cite{ch64a}\cite{ch64b} (see also Harrison et al\cite{wheel2}).

\item The second contribution of Landau is contained in the treatise with Yevgeny Lifshitz, where Landau\cite{lali} gives an estimate of the effects of the electromagnetic interaction between nuclei and electrons and explicitly shows how these interactions can be neglected only in the asymptotic regime for very large densities. The fiducial densities where indeed the Coulomb interactions can be neglected occur at $\rho > 10^7 g/cm^3$ for Z=56, namely at densities too high if compared to the density regime at which the stable white dwarfs and their critical mass occur. The interactions between nuclei and degenerate electrons therefore are not negligible for white dwarfs: they were  certainly difficult to evaluate at the time of the Chandrasekahr's original work.
\end{enumerate}

Actually an  interesting alternative approach, with respect to the evaluation of the interaction term between the electrons and the nuclei, was advanced two years earlier than the work of Chandrasekhar and Landau  by the Soviet physicist Yacov Ilich Frenkel.\cite{FRK} This paper, quoted by Chandrasekhar in his classic book,\cite{chandra} has been in the past and is still today surprisingly ignored. Frenkel proposed to use a relativistic Thomas-Fermi model within a Wigner-Seitz approximation in order to describe stellar matter. The astrophysical motivations were not  clear to Frenkel, though the theoretical formulation he proposed would have deserved a much more thorough examination.

An important turning point came in this field due to the fundamental work of Feynman, Metropolis and Teller. While at los Alamos, they considered the equation of state of compressed matter.\cite{feymete}

It was Ed Salpeter\cite{salpeter,salpeter2}  
who first applied the Frenkel equations and the 
Feynman-Metropolis-Teller\cite{feymete} approach to the treatment of white dwarfs:  Salpeter's  treatment was somewhat affected by the intricacies of the numerical integrations and the drastic approximations he adopted in the numerical solutions.

We ourselves rediscovered the Frenkel Equations and the relativistic Fermi-Thomas treatment applied to an appropriate Wigner-Seitz problem following the study of self-gravitating systems \cite{RB} and the riformulation of the treatment of these systems by using a relativistic generalization of gravitational Fermi-Thomas like equations \cite{ferre} \cite{ruffini} \cite{mrt}. Today with relatively simple numerical computations, we can reconsider the problem and present a comprehensive treatment by first considering the issue of the electromagnetic interactions between nuclei and electrons and turning later to the issue of inverse beta decays.\cite{merlo}  The new treatment\cite{BR} leads to the disappearance of all scaling laws with the chemical composition of the stars, which have been commonly assumed in all treatments up to now. A  new value of the critical mass of white dwarfs is obtained, smaller than the one originally proposed by Chandrasekhar and by Salpeter, and one which strongly depends on the chemical composition of the star. These results appear in principle to be observationally testable. Particularly attractive is the possibility of explaining the observed masses of binary pulsars: these systems are clearly neutron stars which could be formed by the onset of instability of stellar cores very close to the critical mass of white dwarfs as presented by Chandrasekhar.\cite{giacco} The evaluation of their masses will be clearly affected by our computations. 

\section{The Relativistically Compressed Thomas-Fermi Atom}

In the well known Feynman-Metropolis-Teller treatment of the classical Fermi-Thomas model\cite{fermi,thomas} an accurate numerical analysis is made of atomic configurations submitted to external pressure. Following Ferreirinho \textit{et al.} \cite{ferre} and Ruffini and Stella \textit{et al.}\cite{ruffini} we reconsider such a treatment in the framework of a relativistic generalization of the Fermi-Thomas model, which is here briefly derived and solved for different atomic species in different states of compression. Results so obtained are applied to determine the equation of state of cold degenerate stars (white dwarfs) to show how the relativistic and electromagnetic effects acting on microscopic scales affect the overall structure of the star. Finally, the equation of state obtained is used to determine the equilibrium configurations of white dwarfs. A comparison is made with the classic work of Chandrasekhar and Salpeter to show the differences introduced by the use of the new equation of state, and to show how those treatments correspond respectively to a zero-order and first-order approximation to the one we present here. 

Let us consider the spherically symmetric problem of a nucleus with Z protons and A nucleons interacting with a fully degenerate gas of Z electrons. 
The equation of electrostatics for this problem reads, with usual meanings for the symbols appearing in it, 
\begin{equation}
\Delta V(r)=4 \pi e n_e(r)-4 \pi e n_p(r)\ ,
\label{elettro}
\end{equation}
where we have introduced the number density of electrons $n_e(r)$ and of protons $n_p(r)$.

In Eq.~(\ref{elettro}) the quantities $V(r)$ and $n(r)$ are of course not independent, being related by  the equilibrium condition for a relativistic gas in a Coulomb external potential (see ~\cite{lali})
\begin{equation}
c\sqrt{p_{F}^{2}+m^2c^2}-eV(r) = \mbox{const} \equiv E_F \ ,
\label{equili}
\end{equation}
where $E_F$ stands for \emph{Fermi energy}  of the electrons. 

To put the Eq.~(\ref{elettro}) in a simple and dimensionless form we introduce the new function $\Phi(r)$, related to the Coulomb potential by 
\begin{equation} 
\Phi(r)=V(r)+E_F/e
\label{4}
\end{equation}
and the corresponding dimensionless function $\chi(r)$, implicitly defined by 
\begin{equation}
\Phi(r)=\frac{Ze\chi}{r}
\label{6}
\end{equation}
Similarly we have the dimensionless expression for the radius $r=bx$, where
\begin{equation}
b=(3\pi)^{2/3}\frac{\hbar^2}{me^2}\frac{1}{2^{7/3}}\frac{1}{Z^{1/3}} \ .
\label{eq:6}
\end{equation}

We then have
\begin{equation}
p_F=2mc\left(\frac{Z}{Z_{cr}}\right)^{2/3}\left(\frac{\chi}{x}\right)^{1/2}\left[1+ \left(\frac{Z}{Z_{cr}}\right)^{4/3}\frac{\chi}{x} \right]^{1/2} \ ,
\label{impu}
\end{equation}
where
\begin{equation}
Z_{cr}=\left(\frac{3\pi}{4}\right)^{1/2}\left(\frac{\hbar c}{e^2}\right)^{3/2}\approx 2462.4 \ .
\end{equation}
From
\begin{equation}
n_e=\frac{p_{F}^{3}}{3\pi^2\hbar^3}
\label{2}
\end{equation}
we then obtain 
\begin{equation}
n_e=\frac{Z}{4\pi b^3}\left(\frac{\chi}{x}\right)^{3/2}\left[1+ \left(\frac{Z}{Z_{cr}}\right)^{4/3}\frac{\chi}{x} \right]^{3/2} \ .
\label{adens}
\end{equation}

Let us express the second term of the source of Eq.~(\ref{elettro}) in terms of dimensionless quantities: we assume a homogeneous nucleus, with a radius given by the approximate formula 
\begin{equation}
r_{nuc}  = 1.2A^{1/3} \;{\rm Fermi}
\end{equation}
The number density of protons is therefore
\begin{equation}
n_{p}=\frac{3Ze}{4\pi r_{nuc}^3}\Theta\left(x_{nuc}-x \right)
\end{equation}
 
Finally we can write Eq.~(\ref{elettro}) in the form
\begin{equation}
\frac{d^2 \chi}{dx^2}=\frac{\chi^{3/2}}{x^{1/2}} \left[1+ \left(\frac{Z}{Z_{cr}}\right)^{4/3}\frac{\chi}{x} \right]^{3/2}-\frac{3x}{{x_{nuc}}^3}\Theta\left(x_{nuc}-x \right) 
\label{thomfer}
\end{equation}
where $x_{nuc}$ is the dimensionless size of the nucleus ($r_{nuc}=bx_{nuc}$).
Eq.~(\ref{thomfer}) is what we call the ``\emph{generalised dimensionless Fermi-Thomas equation}."

The first boundary condition follows from $ \chi \propto r \Phi $ and therefore $\chi   \stackrel{r{\rightarrow}0}{\longrightarrow}0$, and so
\begin{equation}
\chi(0)=0 \ .
\label{con1}
\end{equation}
The second condition comes from the normalization condition 
\begin{equation}
N=\int_0^{r_0} 4 \pi n_e r^2 dr= Z \int_0^{x_0} \frac{\chi^{3/2}}{x^{1/2}} \left[1+ \left(\frac{Z}{Z_{cr}}\right)^{4/3}\frac{\chi}{x} \right]^{3/2}\;\; x \;\; dx
\end{equation}
with atom size $r_0=bx_0$. Using Eq.~(\ref{thomfer}) we obtain
\begin{equation}
N = Z \int_0^{x_{nuc}} x \chi'' \;\; dx + \frac{3Z}{x_{nuc}^3} \int_0^{x_{nuc}} x^2 \;\; dx + Z \int_{x_{nuc}}^{x_0} x \chi'' \;\; dx
\end{equation}
which gives the simple relation
\begin{equation}
N=Z\left[x_0\chi'(x_0)-\chi(x_0)+1 \right] \ .
\label{con2}
\end{equation}

For a neutral atom we have $N=Z$ and the Eq.~(\ref{con2}) reads simply
\begin{equation}
x_0\chi'(x_0)=\chi(x_0)
\label{con3}
\end{equation}

Note that the physical quantities of interest, such as the Coulomb potential and the density of electrons do not show any singularity in the center, nor on the surface of the nucleus, since they only depend on the function $\chi$ and his first derivative. The only discontinuity appears in the second derivative of $\chi$ due to our rough assumption of a homogeneous nucleus.

\section{The equation of state}

Under these hypothesis, we determine the equation of state of compressed matter simply by  computing for each value of the atomic compression parameter $x_0$ the corresponding  values of pressure and mass density: the first is generated by the degenerate gas of electrons and the second is  given by the mass density of nuclei. 
We assume that the nuclei are arranged in a Wigner-Seitz lattice, each cell being filled by a relativistic gas of degenerate electrons.  The  shape of  the lattice cell is  unimportant in the lowest approximation,\cite{salpeter} so we will assume a spherical cell, with a radius equal to the atomic radius.  The  average density of cells is thus 
\begin{equation}
\overline{\rho}(x_0)= \frac{Am_0}{V_{cell}} = \frac{3 Am_0}{4 \pi (b x_0)^3} \ ,
\end{equation}
where $A$ and $m_0$ are respectively the number and mass of the nucleons.

In this scheme the pressure is also easily determined, being simply the pressure generated by a Fermi gas of number density equal to the number density of electrons at the surface of the atom $n_e(x_0)$, which can be expressed in terms of dimensionless quantities as follows

\begin{equation}
n_e(x_0)=\frac{Z}{4\pi b^3}\left(\frac{\chi(x_0)}{x_0}\right)^{3/2}\left[1+ \left(\frac{Z}{Z_{cr}}\right)^{4/3}\frac{\chi(x_0)}{x_0} \right]^{3/2} \ .
\end{equation}

Our treatment differs from the zero-order approximation of Chandrasekhar (uniformly distributed electrons) and also from the first-order approximation of Salpeter in which the Thomas-Fermi equation for each value of Z and $x_0$ was not solved and  instead an expansion of the number density of electrons was introduced
\begin{equation}
n(x)= n_0 + \epsilon(x)\ .
\end{equation}
This expansion applies only when the electrostatic interaction within nuclei and electrons is small with respect to the Fermi energy of electrons, i.e. in the limit of very high densities.

\begin{figure}[t]
\resizebox{\hsize}{!}{\includegraphics{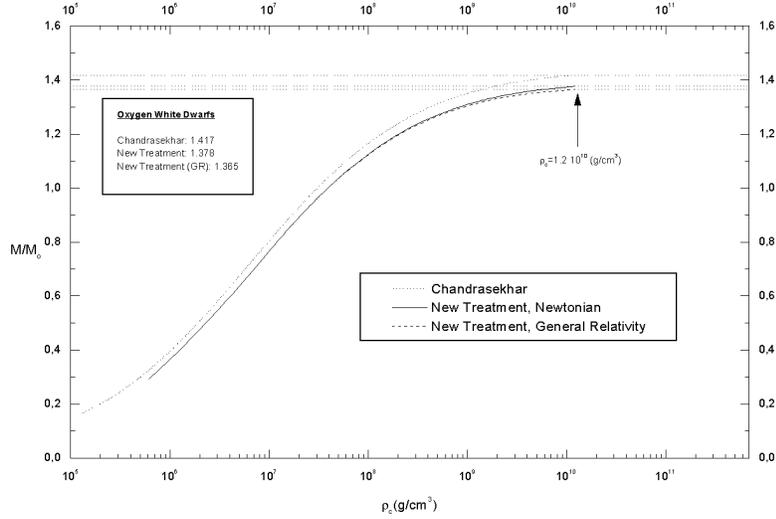}}
\caption[fig]{Equilibrium configurations curve for Oxygen WD in the $M/M_{\odot} -  \rho_c$ plane, obtained in Newtonian theory and General relativity. For comparison we show the results obtained with Chandrasekhar model .Reproduced from Bertone \& Ruffini\cite{BR}.}
\label{ossigeno}
\end{figure}

\section{Equilibrium Configurations for White Dwarfs} 

\begin{figure}[t]
\resizebox{\hsize}{!}{\includegraphics{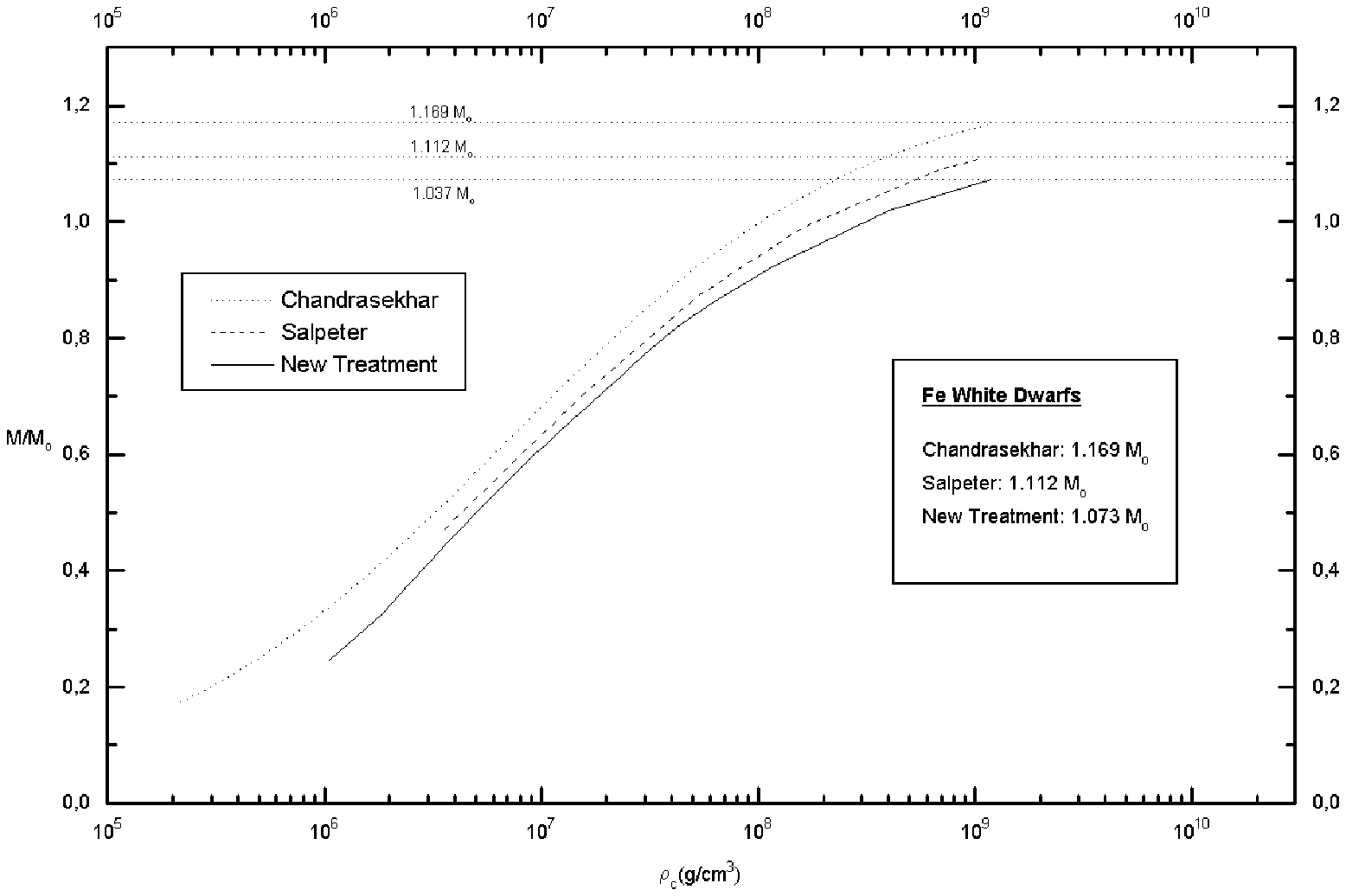}}
\caption[fig]{Equilibrium configurations curve for Iron WD in the $M/M_{\odot} - \rho_c$ plane, obtained in Newtonian theory (the General relativistic curve is pratically superposed at the newtonian one at these low densities). For comparison we show the results obtained by Chandrasekhar and Salpeter .Reproduced from Bertone \& Ruffini\cite{BR}.}
\label{ferro}
\end{figure}

We can now compute the equilibrium configurations for white dwarfs. The results are shown in figures \ref{ossigeno} and \ref{ferro}. We obtain, for the same central density, less massive configurations than those obtained by both Chandrasekhar and Salpeter. Namely for a fixed mass density, i.e. for a fixed value of A and of the compression parameter, our pressure is systematically lower than the one  generated by a uniform shell of electrons.

Numerical integration of the models is performed here for stars of oxygen and iron in the framework of a Newtonian theory of gravitation, as well as in General Relativity. We show that the first one is actually a fairly good approximation up to central densities of the order of $\approx 10^{9}g/cm^3$.

It is shown that deviations from classical models are very important at the lower densities, up to 30\% from the Chandrasekhar model and 10\% from the Salpeter one.

We stop the integration at the critical density corresponding to the onset of inverse beta decays and we thus find the maximum mass for equilibrium configurations. This critical configuration marks also the onset of gravitational instability (see Harrison et al. \cite{wheel2}).

\end{document}